\documentclass{elsart1p}

\usepackage{amsmath}

\usepackage{amssymb}
\usepackage{graphicx}
\newcommand{\be}{\begin{equation}}
\newcommand{\ee}{\end{equation}}
\newcommand{\bea}{\begin{eqnarray}}
\newcommand{\eea}{\end{eqnarray}}

\newcommand{\bm}[1]{\mbox{\boldmath${#1}$}}
 
\newcommand{\uk}{\hat{\mathbf{k}}}

\newcommand{\vs}{\bm{\sigma}}

\begin{document}

\begin{frontmatter}



\title{Supercurrents in color-superconducting quark matter}


\author{Andreas Schmitt}

\address{Institut f\"{u}r Theoretische Physik, Technische Universit\"{a}t Wien, 1040 Vienna, Austria}

\begin{abstract}
We review the basic properties of the currCFL-$K^0$ phase in dense quark matter. At asymptotically large densities,
three-flavor quark matter is in the color-flavor locked (CFL) state. The currCFL-$K^0$ state is a way   
to respond to ``stress'' on the quark Cooper pairing, imposed at more moderate densities by the strange quark mass and the 
conditions of electric and color neutrality. We show how a kaon supercurrent is incorporated in a purely 
fermionic formalism, and show that the net current vanishes due to cancellation of fermion and charge-conjugate fermion
contributions.
\end{abstract}

\end{frontmatter}

\section{Introduction}

Quark matter at large chemical potentials and small temperatures is a color superconductor \cite{Alford:2007xm}. At asymptotically
large chemical potential, the ground state is the color-flavor locked (CFL) phase \cite{Alford:1998mk}. This phase spontaneously 
breaks chiral symmetry, color gauge symmetry and baryon number symmetry. It thus gives rise to Meissner masses for 
all non-Abelian gauge bosons (leaving one combination of the photon and a gluon massless), and there is an octet of 
pseudo-Goldstone bosons, the lightest of which are the 
neutral kaons \cite{Son:1999cm}. In addition, there is an exactly massless superfluid mode. 

It is an important and difficult problem to find the ground state at large, but not asymptotically large, 
densities. In this case, the particularly symmetric CFL state is replaced by a state with less, and less symmetric, Cooper pairing.
Starting from the CFL phase and going down in 
density, one finds that a Bose condensate of kaons develops, the so-called CFL-$K^0$ phase \cite{Bedaque:2001je,Kaplan:2001qk}. 
The next phase down in density seems to be an anisotropic phase where there is a ``supercurrent'' of kaons \cite{Kryjevski:2008zz,Schafer:2005ym},
termed currCFL-$K^0$ phase.

Here we review the basic properties of the currCFL-$K^0$ phase. In particular, we derive the fermionic propagator and the fermionic
dispersion relations. We explicitly show the gauge invariance of the results which allows for a freedom in the choice of the
relative rotation of right- and left-handed gap structures. Although a kaon condensate introduces a difference in right- and left-handed
gap matrices, we shall see that the dispersions of right- and left-handed fermions are identical.
Moreover, we show that the total current in the system vanishes, due to a cancellation
of fermion and charge-conjugate fermion contributions. More ``advanced'' properties of the currCFL-$K^0$ phase, such as the value of the 
current from minimizing the free energy and the calculation of the onset of this phase in terms of the ``stress'' parameter 
$m_s^2/\mu$, where $m_s$ is the strange quarks mass and $\mu$ the quark chemical potential, 
can be found in Refs.\ \cite{Kryjevski:2008zz,Schafer:2005ym,Gerhold:2006np}. 

There are other possible color-superconducting phases which are candidates for the ground state at densities 
reached in compact stars which we shall not discuss here: 
more complicated relatives of the currCFL-$K^0$ phase are crystalline phases \cite{Rajagopal:2006ig}, which contain counter-propagating currents
in more than one spatial direction. If the stress on the pairing becomes too large, the only option is single-flavor pairing, for example
the color-spin locked phase \cite{Schafer:2000tw,Schmitt:2004et}.

\section{Supercurrents in the kaon-condensed phase}

\subsection{Inverse Nambu-Gorkov propagator}

The fermionic part of the Lagrangian, including the condensate of quark Cooper pairs in the mean-field approximation, is
\bea \label{Psi}
\bar{\Psi}S^{-1}\Psi &=& (\bar{\psi},\bar{\psi}_C)
\left(\begin{array}{cc} [G_0^+]^{-1} & \Phi^- \\ \Phi^+ & [G_0^-]^{-1} \end{array}\right)
\left(\begin{array}{c} \psi \\ \psi_C \end{array}\right) \, , 
\eea
where $\Psi$ is a spinor in Nambu-Gorkov space, and where we have neglected the diagonal part of the self-energy  \cite{Wang:2001aq}. 
The inverse tree-level propagator for fermions (+) and charge-conjugate fermions ($-$) 
is given by 
\be \label{G0}
[G_0^\pm]^{-1}=i\gamma^\mu(\partial_\mu - iA^\pm_\mu) \pm \hat{\mu}\gamma^0 \, , 
\ee
with $A_\mu^+\equiv A_\mu$, $A_\mu^-\equiv -A_\mu^T$, and where
the chemical potential $\hat{\mu}$ is a matrix in color-flavor space,
\be \label{mu}
\hat{\mu}\equiv \mu-\frac{\hat{M}^2}{2\mu}+\mu_e Q \, .
\ee
We neglect the masses of the light quarks, 
$m_u\simeq m_d \simeq 0$ and treat the effect of the strange mass as a shift in the chemical potential, i.e., 
$\hat{M}={\rm diag}(0,0,m_s)$ in flavor space. The electric charge chemical potential is denoted by $\mu_e$ and the 
generator of the electromagnetic group is $Q={\rm diag}(2/3,-1/3,-1/3)$ in flavor space. 
The explicit form of the color gauge field $A_\mu$ in Eq.\ (\ref{G0}) shall be discussed below.
For now we only need the Dirac structure of the tree-level propagator which, due to 
our treatment of the quark masses, does not yield mixed terms of left- and right-handed chirality. We are thus left with a sum 
of left- and right-handed inverse propagators
\be \label{Sh}
\bar{\Psi}S^{-1}\Psi=\sum_{h=L,R}\Psi_h^\dag S_h^{-1} \Psi_h \, , 
\ee
where
\be
\Psi_h\equiv \left(\begin{array}{c} \psi_h \\ \psi_{C,h} \end{array}\right) \, , \qquad 
S_h^{-1}\equiv \left(\begin{array}{cc} [G_{0,h}^+]^{-1} & \Phi_h^- \\ \Phi_h^+ & [G_{0,h}^-]^{-1} \end{array}\right) \, , 
\ee
with the two-component spinors $\psi_h\equiv P_h \psi$, $P_h=(1+h\gamma_5)/2$ ($h=+$ for L, $h=-$ for R) and the charge-conjugate counterparts 
$\psi_{C,h}=i\sigma_2\psi_h^*$. The corresponding tree-level propagators are 
\be \label{Gh}
[G^+_{0,h}]^{-1} = P_h\gamma_0[G_{0}^+]^{-1} \, , \qquad  [G^-_{0,h}]^{-1} = -[G_{0}^-]^{-1}P_h\gamma_0 \, .
\ee
In the present calculation it is convenient to use the Dirac matrices in the chiral representation, such that $P_L={\rm diag}({\bf 1},0)$, 
$P_R={\rm diag}(0,{\bf 1})$.

The off-diagonal components of the Nambu-Gorkov propagator contain, as usual for superconductors, the 
condensate of Cooper pairs in form of the gap matrices $\Phi^+=\Delta{\cal M}\gamma_5$, $\Phi^-=-\Delta^*{\cal M}^\dag \gamma_5$,
where $\Delta$ is the gap parameter 
 and where ${\cal M}$ is a $9\times 9$ matrix in color-flavor space. 
We shall assume the gap parameter to be real and to be the same for all color and flavor components (for sufficiently large stress on the pairing
there will be different gaps for different color and flavor components \cite{Gerhold:2006np}). The effect of kaon condensation within 
the color-flavor locked phase 
can be understood as a change in the relative orientation of left- and right-handed gap structures. Therefore, we have different 
gap matrices ${\cal M}_L$ and ${\cal M}_R$,  
\be
\Phi_h^+({\bf x}) = \Delta\,{\cal M}_h({\bf x}) \, , 
\qquad \Phi_h^-({\bf x}) =  - \Delta\, {\cal M}_h^\dag({\bf x})  \, , 
\ee
We consider ${\bf x}$-dependent gap matrices in order to account for
kaon and baryon supercurrents, i.e., nonzero gradients of the Goldstone boson fields. 

\subsection{Supercurrents and gauge fields}

Let us first consider the case without currents and then 
modify the gap structures to include currents. We start from the color-flavor matrices (the subscript ``0'' standing for ``no supercurrent'')
\be \label{M0}
{\cal M}_{0,L} = X_{0,ab} I_a J_b \, , \qquad {\cal M}_{0,R} = Y_{0,ab} I_a J_b \, , 
\ee
(which implies ${\cal M}_{0,L}^\dag = X_{0,ab}^* I_a J_b$, ${\cal M}_{0,R}^\dag = Y_{0,ab}^* I_a J_b$), where 
$(I_a)_{ij}=(J_a)_{ij}=-i\epsilon_{aij}$ $(a,i,j\le 3)$ are antisymmetric matrices in flavor and color space.  
We label the components of the 9-dimensional color-flavor space by the quark modes in the order $ru$, $rd$, $rs$, $gu$, $gd$, $gs$, $bu$, $bd$, 
$bs$, where $r,g,b$ and $u,d,s$ are the colors and flavors, respectively.
The case $X_0=Y_0={\bf 1}$ corresponds to the pure CFL phase, being the ground state at asymptotically large densities. 
If the matrices $X_0$ and $Y_0$ deviate from the unit matrix, the condensate rotates in the space spanned by the generators
which correspond to the broken part of the original symmetry group. Such a rotated order parameter may yield a new ground state
which includes Bose-Einstein condensation of kaons. The value of the kaon condensate $\phi$ is given by 
$\cos\phi=m^2_{K^0}/\mu^2_{K^0}$, where $m_{K^0}$ and $\mu_{K^0}$ are the effective mass and the effective chemical potential of the neutral 
kaon \cite{Kaplan:2001qk}. Condensation occurs for $m^2_{K^0}< \mu^2_{K^0}$ which at moderate densities is likely but not certain, judging from 
extrapolated high-density calculations. For the following we shall assume $m^2_{K^0}\ll \mu^2_{K^0}$, i.e., $\phi\simeq \pi/2$.
For kaon condensation (and none of the other mesons developing a vacuum expectation value), the expectation value of the chiral field 
is
\be \label{sigma0}
\Sigma_0\equiv X_0Y_0^\dag = e^{i\phi T_6} = 
\left(\begin{array}{ccc} 1 & 0 & 0 \\ 0 & \cos\phi & i\sin\phi \\ 0 & i\sin\phi & \cos\phi 
\end{array}\right)\, .
\ee
(Here and in the following, the Gell-Mann matrices are normalized as ${\rm Tr}[T_a T_b]= 2\delta_{ab}$.) 
For a given $\Sigma_0$, $X_0$ and $Y_0$ are only given modulo a color transformation $U\in SU(3)_c$, 
$U(X_0)=X_0U^T$, $U(Y)=Y_0U^T$. One possible choice is
\be \label{xi}
X_0 = Y_0^\dag = \xi \equiv e^{i(\phi/2) T_6} \, .
\ee
Now we add the supercurrents. We allow for nonzero currents $\bm{\jmath}_B$ and $\bm{\jmath}_K$ for the massless bosons
associated with broken baryon number and broken isospin, respectively (the boson associated with superfluidity is 
exactly massless while the Goldstone boson from kaon condensation receives, strictly speaking, a small mass in the 
keV range from weak interactions \cite{Son:2001xd}). 
The kaon current is included by a hypercharge transformation of $\Sigma_0$ \cite{Kryjevski:2008zz,Schafer:2005ym}, 
\be
\Sigma({\bf x}) = V({\bf x})\Sigma_0 V^\dag({\bf x}) \, , 
\ee
with the hypercharge rotation in flavor space
\be \label{hyper}
V({\bf x})=e^{\frac{i}{\sqrt{3}}\theta({\bf x}) T_8}  = \left(\begin{array}{ccc} e^{i\theta({\bf x})/3}& 0 & 0 \\ 0 & 
e^{i\theta({\bf x})/3} & 0
\\ 0 & 0 & e^{-2i\theta({\bf x})/3} \end{array}\right) \, , 
\ee
where $\theta({\bf x})\equiv \bm{\jmath}_K \cdot {\bf x}$. From Eq.\ (\ref{sigma0}) we see that 
this transformation induces a flavor transformation on $X_0$ and $Y_0$,
such that $\Sigma({\bf x})=X({\bf x})Y^\dag({\bf x})$. Any additional color rotation applied on the new 
$X({\bf x})$ and $Y({\bf x})$ should not change the result. We shall show this invariance explicitly by including 
a color rotation $U({\bf x})$,
\be \label{Xcurr}
X({\bf x}) = V({\bf x})\, X_0 \, U^T({\bf x}) \, ,  \qquad Y({\bf x}) = V({\bf x})\, Y_0 \, U^T({\bf x}) \, .
\ee
Furthermore, we add a baryon current whose color-flavor structure is trivial. Consequently, the gap matrices including supercurrents
are given by
\bea \label{MLMR}
{\cal M}_L({\bf x}) &=& e^{-2i\beta({\bf x})}[X({\bf x})]_{ab}I_aJ_b \, , \qquad  
{\cal M}_R({\bf x}) = e^{-2i\beta({\bf x})}[Y({\bf x})]_{ab}I_aJ_b \, , 
\eea
with $\beta({\bf x})\equiv \bm{\jmath}_B \cdot {\bf x}$. We see that the supercurrent state can be viewed as a state with periodically 
varying diquark condensate. The spatial dependence in form of a single plane wave is very ``mild'' since it can be removed by a 
redefinition of the fermion fields (in contrast to multi-plane-wave solutions which result in crystalline structures \cite{Rajagopal:2006ig}). 
To this end, we define the new fields $\varphi_h$ via
\be
\psi_h = e^{i\beta({\bf x})}U({\bf x})V({\bf x}) \varphi_h \, .
\ee
(And thus $\psi_{C,h} = e^{-i\beta}U^*  V^*  \varphi_{C,h}$.) Inserting this transformation into Eq.\ (\ref{Sh}) and 
using the gap matrices (\ref{MLMR}), we obtain the inverse Nambu-Gorkov propagator in the new basis
\bea \label{Sh1}
S_h^{-1}=
\left(\begin{array}{cc} e^{-i\beta}U^\dag V^\dag [G_{0,h}^+]^{-1}UVe^{i\beta}  & -\Delta\, {\cal M}_{0,h}^\dag  \\ \Delta\,{\cal M}_{0,h} & 
e^{i\beta}U^T V^T
[G_{0,h}^-]^{-1}U^*V^*e^{-i\beta} \end{array}\right)
 \, , \qquad 
\eea
where, in the off-diagonal elements, we have used the identities $UJ_b U^T = U^*_{cb} J_c$, $VI_a V^T = V^*_{ca} I_a$.


Next, we specify the gauge field appearing $[G_0^\pm]^{-1}$, see Eq.\ (\ref{G0}). The ansatz which solves the equations of
motion is \cite{Kryjevski:2008zz,Gerhold:2006np,Kryjevski:2004kt}
\begin{subequations} \label{A0A}
\bea
A_0({\bf x}) &=& -\frac{1}{2}[X^\dag({\bf x})\hat{\mu}_0X({\bf x}) + Y^\dag({\bf x}) \hat{\mu}_0Y({\bf x})]^T \, , \\
{\bf A}({\bf x}) &=& -\frac{i}{2}[X^\dag({\bf x})\nabla X({\bf x}) + Y^\dag({\bf x}) \nabla Y({\bf x})]^T \, , 
\eea
\end{subequations}
where $\hat{\mu}_0$ is the traceless part of the chemical potential (\ref{mu}). In terms of the flavor matrices $I_a$, defined below
Eq.\ (\ref{M0}) we have
\be
\hat{\mu}_0 = \left(I_3^2-\frac{2}{3}\right)\frac{m_s^2}{2\mu}+\left(\frac{2}{3}-I_1^2\right)\mu_e \, .
\ee
While $\hat{\mu}_0$ is a matrix in flavor space, the matrix products $X^\dag\hat{\mu}_0X$ and $Y^\dag\hat{\mu}_0Y$ are color matrices
since $X$ and $Y$ 
have one color and one flavor index as can be seen from Eq.\ (\ref{MLMR}). This conversion from flavor to color space via a usual 
matrix multiplication is a consequence of color-flavor locking.  
We can now insert the explicit forms of $X({\bf x})$ and $Y({\bf x})$ from Eq.\ (\ref{Xcurr}) as well as the matrices $X_0$ and $Y_0$ 
from Eq.\ (\ref{xi}) with the condensate
$\phi=\pi/2$ into the gauge fields (\ref{A0A}). As a result, the only remaining ${\bf x}$-dependence is left in the color transformations
$U({\bf x})$, and we find 
\begin{subequations} \label{gauge1}
\bea
A_0({\bf x}) &=& U({\bf x})\left(\frac{m_s^2}{4\mu}+\mu_e\right)\left(J_1^2-\frac{2}{3}\right)U^\dag({\bf x}) \, , \\
{\bf A}({\bf x}) &=& U({\bf x})\frac{\bm{\jmath}_K}{2}\left(\frac{2}{3}-J_1^2\right)U^\dag({\bf x}) -i[\nabla U({\bf x})]U^\dag({\bf x}) \, .
\label{gauge12}
\eea
\end{subequations}
This result shows that the gauge
fields transform under a color transformation as expected, $A_0\to UA_0 U^\dag$, ${\bf A}\to U{\bf A}U^\dag - i(\nabla U)U^\dag$.

\subsection{Gauge invariance and inversion of propagator}

Now we insert the gauge fields (\ref{gauge1}) into the inverse propagator (\ref{Sh1}), wherefore we use Eqs.\ (\ref{G0}) and (\ref{Gh}).
The result is (in momentum space) 
\bea \label{Sh2}
S_h^{-1}=
\left(\begin{array}{cc} k_0+\mu_{\rm eff}-h\bm{\sigma}\cdot({\bf k}+{\bf A}_{\rm eff})   & -\Delta\, {\cal M}_{0,h}^\dag  \\ 
\Delta\,{\cal M}_{0,h} & -[ k_0-\mu_{\rm eff}+h\bm{\sigma}\cdot({\bf k}-{\bf A}_{\rm eff})] \end{array}\right)
 \, ,  
\eea
with the color-flavor matrices
\bea
\mu_{\rm eff} &\equiv& \hat{\mu}+\left(\frac{m_s^2}{4\mu}+\mu_e\right)\left(J_1^2-\frac{2}{3}\right) \, , \qquad 
{\bf A}_{\rm eff} \equiv -\bm{\jmath}_B+\bm{\jmath}_K\left(1-\frac{1}{2}J_1^2-I_3^2\right) \, .
\eea
We have thus explicitly shown that the gauge transformation $U({\bf x})$ drops out of the result, as expected. 
This shows, in particular, that we could have used any other 
form of $X_0$, $Y_0$, related by a color transformation to our choice (\ref{xi}); this would simply have added a constant 
color rotation to the $U({\bf x})$ from above.

Let us denote the entries of the propagator $S_h$, i.e., the inversion of Eq.\ (\ref{Sh2}), by
\be
S_h = \left(\begin{array}{cc} G_h^+ & F_h^- \\ F_h^+ & G_h^- \end{array}\right) \, ,
\ee
where $F_h^\pm$ are the so-called anomalous propagators. From Eq.\ (\ref{Sh2}) one
finds
\begin{subequations}
\bea
\hspace{-0.51cm} G_h^\pm &=& \pm \left[k_0 \pm \mu_{\rm eff} \mp h\vs\cdot ({\bf k}\pm {\bf A}_{\rm eff})-\Delta^2 {\cal M}_{0,h}^\mp
\frac{k_0\mp \mu_{\rm eff} \mp h \vs\cdot ({\bf k}\mp {\bf A}_{\rm eff})}{(k_0\mp\mu_{\rm eff})^2-({\bf k}\mp {\bf A}_{\rm eff})^2}\,{\cal M}^\pm_{0,h}
\right]^{-1}  \\
\hspace{-0.5cm} F_h^\pm &=& -\Delta \, G_{0,h}^\mp {\cal M}_{0,h}^\pm G_h^\pm \, , 
\eea
\end{subequations}
where we have denoted ${\cal M}_{0,h}^+\equiv {\cal M}_{0,h}$, ${\cal M}_{0,h}^-\equiv {\cal M}_{0,h}^\dag$. The remaining inversion is very 
complicated due to the nontrivial entanglement of Dirac and color-flavor space. The situation simplifies considerably if
we neglect the transverse part of the gauge field, ${\bf A}_{\rm eff} \simeq \uk\,{\bf A}_{\rm eff}\cdot \uk$, where 
$\uk\equiv {\bf k}/|{\bf k}|$. This is a good approximation for small currents. In this
case we can write
\begin{subequations} \label{GF}
\bea
G_h^\pm &=& \pm\sum_{e}\lambda_{\bf k}^{e,h}\left[Z_e^\pm -\Delta^2{\cal M}_{0,h}^\mp (Z_e^\mp)^{-1}{\cal M}_{0,h}^\pm
\right]^{-1} \, , \label{Gh5}\\
F^\pm_h &=& \pm\Delta\sum_e \lambda_{\bf k}^{e,h}(Z_e^\mp)^{-1}{\cal M}_h^\pm
\left[Z_e^\pm -\Delta^2{\cal M}_{0,h}^\mp (Z_e^\mp)^{-1}{\cal M}_{0,h}^\pm
\right]^{-1} \, , \label{GF2}
\eea
\end{subequations}
with the orthogonal projectors $\lambda_{\bf k}^{e,h}\equiv (1+eh\vs\cdot\uk)/2$ and 
\be \label{Zecurrent}
Z_e^{\pm}=k_0\pm(\hat{\mu}_{\rm eff}-ek)-e{\bf A}_{\rm eff}\cdot\uk \, .
\ee
Now the inversion in color-flavor space has decoupled: the matrices in the 
square brackets in Eqs.\ (\ref{GF}) are pure color-flavor matrices.
 
\section{Charge density and vanishing net current}

In order to understand the physics of the currCFL-$K^0$ phase it is instructive to compute the electric charge density, written
as a sum of the fermion and charge-conjugate fermion contributions as well as contributions from the left- and right-handed 
fermions,
\be \label{nQsum}
n_Q = \sum_h(n_{Q,h}^+ - n_{Q,h}^-) \, , \qquad n_{Q,h}^\pm\equiv {\rm Tr}[QG_h^\pm] \, ,
\ee
where the trace is taken over color-flavor, Dirac, and momentum space. 
Neglecting the antiparticle contribution, denoting the elements of the diagonal matrix (\ref{Zecurrent}) by
$Z_{e=+}^\pm={\rm diag}(z_1^\pm, \ldots ,z_9^\pm)$, and using $z_4^\pm=z_7^\pm$ (which follows from the explicit forms of
$\mu_{\rm eff}$ and ${\bf A}_{\rm eff}$), we find after performing the inversion in Eq.\ (\ref{Gh5}) and the color-flavor trace
in Eq.\ (\ref{nQsum})
\bea \label{nQpm}
\hspace{-0.5cm}
n_{Q,h}^\pm &=& \frac{2}{3}\frac{T}{V}\sum_K\left(2\frac{z_2^\mp}{z_2^\mp z_4^\pm -\Delta^2}-\frac{z_4^\mp}{z_2^\pm z_4^\mp -\Delta^2}
+2\frac{z_3^\mp}{z_3^\mp z_4^\pm -\Delta^2}-\frac{z_4^\mp}{z_3^\pm z_4^\mp -\Delta^2}\right)+ \ldots 
\eea
We observe that the result is independent of $h$, i.e., the same for left- and right-handed contributions. The reason is that the color flavor part 
of $G_R^\pm$ is simply the transpose of the color-flavor part of $G_L^\pm$. This means in particular that left- and right-handed
fermions have the same dispersion relation.
  
Here we are not interested in the exact result of the charge density but rather want to explain the general features
of the phase with a supercurrent. We have thus omitted the complicated terms coming from a $5\times 5$ block in the propagator.
The remaining contribution shown in Eq.\ (\ref{nQpm}) has contributions from two Cooper pairs.
The first two terms on the right-hand side give the contribution of pairing of 
a red down quark with a mixture of green and blue up quarks, i.e., in terms of flavors this is a down-up pair. 
In the second contribution the $rd$ quark is replaced by an $rs$ quark, this is a strange-up pair. In both cases, 
one Cooper pair constituent has twice the negative charge of the other, hence the factors of 2 and the minus signs. 
For illustrative purposes, let us pick one of the pairs, say the strange-up pair, and define the occupation numbers 
\be \label{occ}
f_{1,2}^+({\bf k})\equiv T\sum_{k_0} \frac{z_{4,3}^-}{z_3^\pm z_4^\mp -\Delta^2} \, , \qquad 
f_{1,2}^-({\bf k})\equiv T\sum_{k_0} \frac{z_{4,3}^+}{z_3^\mp z_4^\pm -\Delta^2}\, ,
\ee 
such that the electric charge density becomes
\be \label{nQ2}
n_Q = \frac{4}{3}\int\frac{d^3{\bf k}}{(2\pi)^3}\left\{2[f_2^+({\bf k})-f_2^-({\bf k})]-[f_1^+({\bf k})-f_1^-({\bf k})]\right\} + \ldots 
\ee
The occupation numbers $f_{1,2}^+$ are positive while the occupation numbers for the charge-conjugate fermions $f_{1,2}^-$ are negative.
The subscripts 1 and 2 correspond to the two constituents of the Cooper pair. In the conventional BCS pairing without 
supercurrent, \mbox{$f_1^+=f_2^+ = -f_1^-=-f_2^-$}. 
Let us now define
\bea
\xi_\pm&\equiv& k-\left(\bar{\mu}\pm\frac{3\bm{\jmath}_K}{4}\cdot \uk\right) \, , \qquad 
\epsilon_\pm \equiv  \sqrt{\xi_\pm^2+\Delta^2} \, , \qquad 
\jmath\equiv  \left(\bm{\jmath}_B-\frac{\bm{\jmath}_K}{4}\right)\cdot\uk \, ,
\eea
where $\bar{\mu}$ is the ``common'' chemical potential of the two constituents at which pairing takes place, in this specific case 
$\bar{\mu}=\mu-7m_s^2/(24\mu)$. Moreover, we denote the particle ($p$) and hole ($h$) dispersions, which are the poles of the 
propagator, and thus also the poles in Eq.\ (\ref{occ}), by
\bea
E_p^\pm \equiv \epsilon_\mp\pm\delta\mu-\jmath \, , \qquad 
E_h^\pm \equiv -\epsilon_\mp\pm\delta\mu-\jmath \, ,
\eea
where $\delta\mu$ is the ``mismatch'' in chemical potentials, in this case $\delta\mu=\mu_e+m_s^2/(8\mu)$. We can 
now perform the Matsubara sum in Eq.\ (\ref{occ}) and obtain, after subtracting the vacuum contribution,
\begin{subequations}
\bea
f_{1/2}^+({\bf k}) &=& -\frac{1}{2}\left(\frac{\epsilon_\mp+\xi_\mp}{2\epsilon_\mp} \tanh\frac{E_p^\pm}{2T}+ 
\frac{\epsilon_\mp-\xi_\mp}{2\epsilon_\mp} \tanh\frac{E_h^\pm}{2T}-1\right) \, , \\
f_{1/2}^-({\bf k}) &=& -\frac{1}{2}\left(\frac{\epsilon_\pm+\xi_\pm}{2\epsilon_\pm} \tanh\frac{E_h^\mp}{2T}+ 
\frac{\epsilon_\pm-\xi_\pm}{2\epsilon_\pm} \tanh\frac{E_p^\mp}{2T}+1\right) \, . 
\eea
\end{subequations}
Let us assume that $\bm{\jmath}_B=\jmath_B\hat{\bf z}$ and $\bm{\jmath}_K=-\jmath_K\hat{\bf z}$, such that 
$\jmath=(\jmath_B+\jmath_K/4)\, x$, where $x=\cos\theta$ and $\theta$ is the angle between ${\bf k}$ and the $z$-axis. 
This implies $f_{1/2}^+(x) = -f_{1/2}^-(-x)$. The angular integration 
in the charge density (\ref{nQ2}),
\be
\int_{-1}^1 dx\,f_{1/2}^+(x) =-\int_{-1}^1 dx\,f_{1/2}^-(x) \, ,
\ee
thus ensures that fermions and charge-conjugate fermions yield the same contribution to $n_Q$ (note the additional minus in Eq.\ (\ref{nQ2})
in front of the charge-conjugate terms). We also have the relation   
\be \label{fx}
\int_{-1}^1 dx\,x f_{1/2}^+(x) =\int_{-1}^1 dx\,x f_{1/2}^-(x) \,  .
\ee
This relation ensures that the total electric current vanishes,
\be
{\bf J}_Q = \frac{4}{3}\int\frac{d^3{\bf k}}{(2\pi)^3}\,{\bf k}\,
\left\{2[f_2^+({\bf k})-f_2^-({\bf k})]-[f_1^+({\bf k})-f_1^-({\bf k})]\right\} + \ldots =0 \, .
\ee
In the directions perpendicular to $\hat{\bf z}$, ${\bf J}_Q$ vanishes trivially; in the $z$-direction it vanishes 
since fermion and charge-conjugate fermion contributions cancel each other, as can be seen from Eq.\ (\ref{fx}). 
This implies that the net current corresponding to any conserved
charge must vanish. Here we have shown this on a purely fermionic level, adding up the contributions of all ungapped fermions. 
Another way of saying this is that the gradient of the Goldstone boson field is cancelled by a current of ungapped fermions 
\cite{Schafer:2005ym,Son,Son:2007ny}.   

\section*{Acknowledgments}

I thank the organizers of the SEWM 2008 conference in Amsterdam for a stimulating and fruitful conference. 
I thank Mark Alford and Andrei Kryjevski for valuable discussions
about the work reviewed here. I acknowledge support by the FWF project P19526.

\end{document}